# Stability of different phases of $(C_{60})_2$ Structures

Narinder Kaur, Shuchi Gupta, Keya Dharamvir and V. K. Jindal

Department of Physics, Panjab University, Chandigarh -160014, India

**ABSTRACT**

We investigate the possible binding configurations of pairs of $C_{60}$ molecules when pushed against each other. Tersoff potential, which represents intramolecular interactions well, has been used to calculate potential energies. We begin relaxation of atomic coordinates at various distances of separation and for all possible mutual orientations of the two molecules. As a result, we have been able to show that several minimum energy configurations exist. Some of these configurations have not been reported earlier. Only two types of dimer structures, involving interlinkage through a single bond, or through so called 2+2 cycloaddition, have been commonly referred in the literature. Our calculation shows that apart from these configurations, many interesting composite phases also result, such as fused and peanut structures and (5,5) and (10,0) nanotubes. A link with experiment to find these structures can be established by application of suitable critical applied pressure in the solid phase, accompanied by high temperature corresponding to orientational melting so that suitable mutual orientations are available. High energy molecular beams of $C_{60}$ incident upon $C_{60}$ layers could also achieve the same.

# 1. INTRODUCTION

The 60-atom carbon clusters with truncated icosahedral structure, popularly known as buckyballs, have been intensively investigated for the past two decades. They form molecular crystals with weak intermolecular bonding, adequately represented by Van der Waals interactions (see, e.g., Ecklund and Rao, 2000). In this crystalline state, at ambient temperature, the buckyballs are free to rotate around the molecular centers while preserving a perfect crystalline lattice order. At temperatures below 350K the orientational freedom freezes.

A gentle push, provided by hydrostatic pressure, excitation by light or other factors can promote a stronger covalent bonding between the $C_{60}$ molecules, thus allowing them to share some of their electrons. This process leads to formation of dimers, polymer like chains or 2 and 3 dimensional rigid networks (Rao et. al., 1993, Xu and Scuseria, 1995). This may dramatically change the electronic and optical properties of the bulk material and displays numerous fascinating properties of a collective nature. In these new phases, equilibrium distances between nearest neighbour $C_{60}$ molecules get shortened from 9.9Å (in the crystalline state) to about 9.0Å. The resulting solids can have orthorhombic, tetragonal or monoclinic structures. One obtains a dimer phase when $C_{60}$ soild is cooled rapidly from 450K to 77K and subsequently heated to 200K (Schober and Renker, 1999). Similarly, chain and layered polymer phases have been produced by cooling the $C_{60}$ solid slowly from 450K to 77K. The phase purity has been checked by X-Ray diffraction. The dimer phase is metastable and changes gradually to chain polymer phase. The bond between two adjacent buckyball monomers in these structures is either a single bond (Oszlanyi et. al., 1996) or a 2+2 cycloaddition bond(Adams et. al., 1994 and Menon et. al., 1994). The center to center distance between two buckyballs in a single-bond-type dimer is 9.1Å while that in a cyclo-added dimer is 9.3Å. Iijima and his coworkers heated nano-peapods (rows of buckyballs inside single walled carbon nanotubes) and showed that inside the nanotube, the buckyballs coalesce at $800^0C$ -$1200^0C$ (Rueff et. al., 2002).

A number of theoretical studies have also been made to investigate $C_{60}$ dimerization. The binding energies of the dimer are estimated to be 1.20eV per single bond (Porezag et. al., 1995) and 1.25 eV for the cycloaddition bond (Adams et. al., 1994). There is also evidence of some theoretical work on peanut (Andriotis et. al., 2003) and bucky-tube formation (Ueno *et al.* 1998).

While studying fullerene isomerization a particular atomic rearrangement called Stone-Wales (SW) or pyracyclene transformation, i.e., a 90° bond rotation within the plane of a $sp^2$-carbon network was found to have a key role (Zhao et. al., 2003). Ueno *et al.* (1998) first reported a complete, ''seamless'' fusion of two buckyballs through a series of SW rotations

In the collision approach (Xia et. al., 1997), using molecular dynamics simulations it has been shown that dumbbell-shaped dimers with almost intact cages are formed at low collision energies whereas at high enough energies the fusion barrier is overcome and the two colliding $C_{60}$ molecules fuse to form one large cage-cluster.

Solids formed out of dimerized $C_{60}$ have also been studied theoretically (Kaur, N., et. al., 2000, Dzyabchenko et.al., 1999). Some of the thermal, phonon related properties have been calculated by us. A set of dimer lattices have been derived by Dzyabchenko et.al (1999). Such studies assume the dimer to be formed through either single bond or cycloaddition.

In this paper, we investigate the possible stable structures of the dimer molecule, when two buckyballs are brought close to each other at a defined orientation. In this way we have obtained ten structures, including the well known single and cyclo-added dimers. The numerical results of these dimer structures obtained are presented and discussed. The required pressure and temperature conditions to obtain various dimer structures, on compressing solid $C_{60}$, have also been calculated. Required velocity of a beam of buckyballs for likely production of a certain type of dimer when impinging on a surface of bucky solid, has also been calculated.

## 2. THEORETICAL MODEL AND CALCULATIONS

We have used a theoretical model in which the interaction between bonded carbon atoms is governed by Tersoff's potential (1988 a). We calculate binding energies of two buckyball systems at various intercage distances for all possible orientations. We simulate this situation by first placing the $C_{60}$ molecules (i.e. their coordinates) at a short distance apart, in such a way that they are within covalent bonding range of each other (≈2.1Å). Then we allow the coordinates of all 120 atoms to relax in such a way that eventually a minimum in energy is reached. The new structure thus obtained is analyzed to obtain the number of bonds, bond energies, lengths and other characteristics.

### A. *The model Potential*

The potential consists of a pair of Morse-type exponential function,

$$V_{ij} = f_c(r_{ij})[e^{-\lambda_1 r_{ij}} - b_{ij} B e^{-\lambda_2 r_{ij}}] \tag{1}$$

the two terms within square brackets describing the repulsive and attractive parts respectively. $f_c(r)$ is a cut-off function which varies from 1 to 0 in sine form between R-D and R+D, D being a short distance around the range R of the potential,

$$f_c(r) = \begin{cases} 1, & r < R - D \\ \frac{1}{2} - \frac{1}{2}\sin\left[\frac{\pi}{2}(r-R)/D\right], & R - D < r < R + D \\ 0, & r > R + D \end{cases} \tag{2}$$

The other functions in equation 1 are,

$$b_{ij} = \frac{1}{\left(1 + \beta^n \xi_{ij}^n\right)^{\frac{1}{2n}}}, \tag{3}$$

where, $\xi_{ij} = \sum_{k \neq i,j} f_c(r_{ik}) g(\theta_{ijk}) e^{\left[\lambda_3^3 (r_{ij} - r_{ik})^3\right]}$. (4)

Here $\theta_{ijk}$ is the bond angle between ij and ik bonds.

$$g(\theta) = 1 + \frac{c^2}{d^2} - \frac{c^2}{\left[d^2 + (h - \cos\theta)^2\right]}. \tag{5}$$

Using this potential, composite energy of all the atoms of the system is given by $E_b$ which is written as

$$E_b = \frac{1}{2} \sum_{ij} V_{ij} \tag{6}$$

The sum in Eq. 6 includes all the atoms in each of the molecules, running from 1 to 120. The Van der Waals interaction potential operative for interactions between the non bonded i$^{th}$ and j$^{th}$ carbon atoms is numerically insignificant at the distances of consideration, compared to the Tersoff potential, and has not been included in the present calculation.

The Tersoff potential has been successfully used for the modeling of chemical bonding in a wide range of hydrocarbon molecules, diamond, graphite and carbon nanotubes and is able to distinguish among different carbon environments, fourfold $sp^3$ bond as well as threefold $sp^2$ bond. The parameters of this potential are presented in Table I, where the first four differ from Tersoff's as they were adjusted to produce better agreement with measured values of $C_{60}$ bond lengths and energies. The parameters $\lambda_3$ is usually taken as 0. However the corresponding factor in eq. 4 may contribute significantly for atoms outside the first neighbor shell. We take $\lambda_3$ equal to $\lambda_2$, rather than 0, so that a buckyball under compression gives physically valid results.

### B. *Obtaining the Dimer Structure*

We adopt a static procedure (relaxation) where the initial configuration consists of two buckyballs situated a certain distance apart (~8.5Å center to center) with a certain mutual orientation. The atomic coordinates of the 120 atoms are then adjusted one by one to obtain a configuration with lower energy (eq.6). The cycle is repeated many times till reasonably stable energy (viz. the minimum) is obtained.

We identify 15 different mutual orientations of two $C_{60}$ molecules which lead to distinct dimer structures. Various combinations of single bond (SB), double bond (DB), corner (C) atom, hexagonal face (HF) and pentagonal face (PF) which come face to face with each other in the two buckyballs, make different starting configurations.

For each orientation, starting intercage distance was taken as that at which the energy of the unrelaxed composite $C_{120}$ molecule is minimum. The plot for one such orientation (DB-DB) is shown in fig. 1. The flexibility of the $C_{60}$ molecule is now introduced allowing the movement of carbon atoms, to get minimized structure. For some of the resulting structures, such as the $C_{120}$ nanotubes, this procedure does not work. For these, the initial configuration must consist of distorted $C_{60}$ molecules (some of the on-cage bonds already open). Within our treatment, this is the only way to arrive at the final states of carbon nanotube $C_{120}$ and the peanut. The cage-opening represents thermal activation as has also been described by Marcos et. al. (1997).

### 3. RESULTS

### A. *Dimer Structures obtained*

In fig.1, the first minimum is at intercage distance 9.1Å which give rise to structure 5 after relaxation. The second minimum at intercage distance 8.6Å gives rise to structure 1 in fig. 2. Similar plots for other possible orientations were studied and it was found that initial intercage distances of 8.0Å to 9.0Å usually yield bonded structures. The dimers obtained were categorized depending upon their bonding schemes and are shown in Fig 2. we find that the possible three classes are a) dumbbell structures -- with very few bonds, not significantily disturbing carbon atoms other than those involved in intercage bonding; b) Fused structures – those in which contact atoms have some of the original $C_{60}$ bonds broken and new bonds formed- mixture of $sp^2$-$sp^3$; c) Coaslesced structures -- those with all of the carbon atoms finally attaining $sp^2$. Further features of these structures are discussed in the next section. The numerical results are summarized in Table-II.

We define the center to center distance, as the distance between the center of gravity of the first 60 atoms and that of the last 60 atoms, originally belonging to the two buckyballs and dimer length is defined as the end to end axial distance between the two balls as shown in fig. 3. In reality only a few atoms, the "contact atoms" relax appreciably. The structures obtained after relaxation from open cages require, in addition to initial proximity, some initial extra energy, which could be provided by temperature.

The fused and some single bonded dimer structures discussed here are the new structures which have scanty reference in literature. There is no reason why these structures cannot be obtained experimentally. We have not considered the energetically stable, nearly icosahedral, $C_{120}$ cages studied by Esfarjani et.al (1998) or toroidal cage form studied by Ihara et.al. (1993), because our procedure of obtaining a $C_{120}$ structure was by compressing two $C_{60}$ monomers such that the two balls retain their individuality at least by 50%, after the dimer formation whereas for the above mentioned structures the two balls completely lose their identities.

The results of our calculations are shown in fig.2 and Table II. Below, we discuss these obtained structures.

a) **Single bonded and cycloadded dimers, the dumbbells**

The dimer structures under this category are formed when the initial intercage distances are from 8 to 9Å. For different orientations, different bonding schemes result. In **structure 1** bonding is through cycloaddition, of bond length 1.54 Å each, whereas the intramolecular bonds of this ring are 1.47 Å each. This type of bonding is the much talked about bonding in the $C_{60}$ dimers as well as chain polymers. The central $C_4$ unit connecting the buckyballs can be viewed as a cyclobutane fragment; every carbon atom is connected to four others. **Structure 2** is more stable than structure 1 although the distortion at the interconnecting sites is similar. These two structures are the most commonly (experimentally) referred dimer structures. **Structures 3, 4 and 5** are all singly connected but have different symmetries. Structures 2 and 5 look similar but have different energies and initial orientations. In fact 2, with lower energy has shorter route during relaxation. Given enough time, structure 5 also relaxes to 2.

b) **Fused dimers**

The dimer structures under this category are formed when the initial intercage distances are between 8.0 and 8.5 Å. **Structure 6** was obtained from SB-DB whereas **Structure 7** was obtained from SB-SB configuration. An intramolecular bond of one ball breaks and two intermolecular bonds form giving rise to $sp^2$ like bonding at the interconnecting site. This type of structure was also obtained by Choi et. al. (1998) theoretically.

**c) Coalesced dimers**

The dimer structures under this category were formed with initial intercage distances less than 8.2 Å. "Peanut"(**structure 8**) was obtained by pushing one partially opened ball shown in fig 4 towards closed HF of the second ball. The resulting structures show all C-atoms, near the interlinking site to be $sp^2$ bonded. The "Coalesced" structure has been confirmed by laser desorption mass spectroscopy and its structure was assigned a peanut shape by comparison of the IR absorption spectrum with theoretical spectra of five $C_{120}$ isomers by Strout et al (1993). Two more peanut structures have been studied by Hara and Onoe (2003). These structures have been observed by Kim et. al. (2003) in the electron beam-irradiated $C_{60}$ thin film as well as in the photo-irradiated $K_xC_{60}$ film. Hara et. al. (2000) also found the coalesced dimers in aggregation following laser ablation of fullerene films, in collision between fullerene ions and thin films of fullerenes and in fullerene- fullerene collisions. The Buckyballs have also been observed to similarly coalesce inside a peapod.

**Structure 9,** the $C_{120}$ molecule in the form of armchair Buckytube has been obtained by bringing together two $C_{60}$ monomers approaching each other in such a way that partially open pentagons are facing each other. This 'partial opening' has been shown in fig 5. The contacting pentagon has all its single bonds cut so that all five C-atoms move away from each other; subsequently one bond from each of the next layer of pentagons is also cut. As the two cages are now open they can fit into each other if brought very close, resulting in a nanocapsule of length 11.84Å. Theoretically, breaking of the ten bonds of each of the two $C_{60}$ molecules seems an easy way to open the cages, but it is believed that the isomerization mechanism is preferred as there is less expenditure of energy for the SW transformations. To quote Onoe et. al. (1999), a sequence of only five SW-type bond rotations transforms a perfect $C_{60}$ molecule to a capped segment of a (5,5) nanotube. **Structure 10,** the *Zigzag Buckytube* was not attainable by any kind of cage opening as a precursor. Instead, we fed the assumed structure to the program and let it minimize. The resulting zigzag tube has a length of 12.30 Å, and is the most stable of the dimerized $C_{60}$ molecules found by us.

Fullerene coalescence, experimentally found in fullerene embedded SWNTs under heat treatment, has been simulated by Kim et.al (2003) who took the initial state as two $C_{60}$ molecules separated by 1nm. The synthesized inner tubes had their diameters ranging from 0.6-0.9 nm. Esfarjani et. al. (1998) performed total energy minimizations for structures 1, 3, 7 and 9.

**B.     The $C_{60}$ solid at high pressure and temperature -- starting point for dimerisation**

In the computer simulation, we have considered only two buckyballs which are allowed to relax after having been brought close together at a certain mutual orientation. If we are to have the same conditions in a solid, we must apply high pressure to bring the buckyballs close. However, at ambient temperature, the balls are frozen in mutually fixed orientation, viz., DB-HF.  Therefore, the temperature must be raised upto that of orientational melting, so that overcoming the orientational potential barrier, all mutual orientations (including the required one) become possible. Further, at high temperatures, the buckyballs execute large amplitude oscillations (phonons) thus bringing temporarily closer any two nearest neighbour buckyballs.

The mean square displacement $<u_0^2>$ of the molecules at temperature $T$ and Debye frequency $\omega_D$, are related through equation 7 (see, e.g., Kittel, 1996). Due to thermal vibrations the molecules come closer periodically, if there is an energy trap at some inter-cage distance, then the $C_{60}$ molecules are likely to stay there and form dimer bonds.

$$\langle u_0^2 \rangle = \frac{3\hbar}{\rho V \omega_D} \coth\left(\frac{\hbar \omega_D}{2 k_B T}\right) \tag{7}$$

$$\omega_D^3 = 6\pi^2 v_S \frac{N}{V} \tag{8}$$

where, $N/V$ is the molecular density, $k_B$ Boltzmann constant, $T$ absolute temperature, $v_S$ velocity of sound in the solid and $\rho$ is its density. The orientational melting temperature $T_m$ at different pressures has been estimated by making use of our earlier calculation of orientational barrier to the spinning of a single $C_{60}$ molecule in the solid (Dharamvir and Jindal, 1992).

Table III lists a set of possible P-T values which provide the initial conditions (Column 4 of Table II). The temperatures in Table III are small compared to bond energies of $C_{60}$ (~3-5eV), which justifies our model -- we let the two buckyballs come close to begin with, and then allow relaxation of C atoms within the molecule. The last column of this table lists buckyball velocities equal to the initial potential energies quoted in Table II so that beams of these initial energies could provide adequate conditions for dimerisation to a particular phase.

## 4. SUMMARY AND CONCLUSION

We study the various forms of dimers of $C_{60}$ obtained after squeezing together two buckyballs. Table II shows that the most stable dimer structures are the ones that started (before relaxation) with highest energies. This indicates that these are the structures which have to overcome the highest potential barriers.

Comparing all these structures we find that the most stable dimers in our work were coalesced ones obtained when partially open monomers or isomers of $C_{60}$ molecules with PFPF or HFHF orientation, were brought to a distance less than 8.1Å and allowed to relax. However, if unopened balls are given the same initial configuration then, they tend to fuse together by breaking one or two intramolecular bonds and forming multiple intermolecular bonds as in structure 7. Sometimes the C-atoms occupy position in between the buckyballs thus losing their initial identity. For inter-cage distances between 8Å and 9Å, the facing bonds break and dumbbell structures are formed as the balls retaining their individuality.

We have also suggested that the required preconditions could be simulated by compressing the $C_{60}$ solid. High pressure brings the $C_{60}$ molecules within chemical bonding range and high temperature causes the orientational repositioning. After freezing the system (i.e., quenching), the intramolecular interaction potential is switched on to facilitate the relaxation of all the 120 atoms to achieve minimum energy configuration.

The results and inferences of this work provide motivation for experimentation on the $C_{60}$ dimer molecule forming the dimer solids. For the discussed dimer structures we propose to investigate the consequent structures of the dimer solids, in line with our earlier work (2000).

## Figure Captions

**Figure 1:** Energy (according to eq.6) of two rigid buckyballs at a fixed orientation and varying intercage distances. In this orientation, one double bond on each ball faces another one on the other ball in a parallel manner.

**Figure 2:** Various structures obtained after relaxation under Tersoff potential. The groups A, B and C refer to "dumbbells", "fused" and "coalesced" structures respectively.

**Figure 3**: Showing intercage distance and dimer length.

**Figure 4**: Front and Side view of opened cage for peanut structure.

**Figure 5:** The seamless joining of two partially open $C_{60}$ molecules to form an armchair buckytube.

**Figure 1**

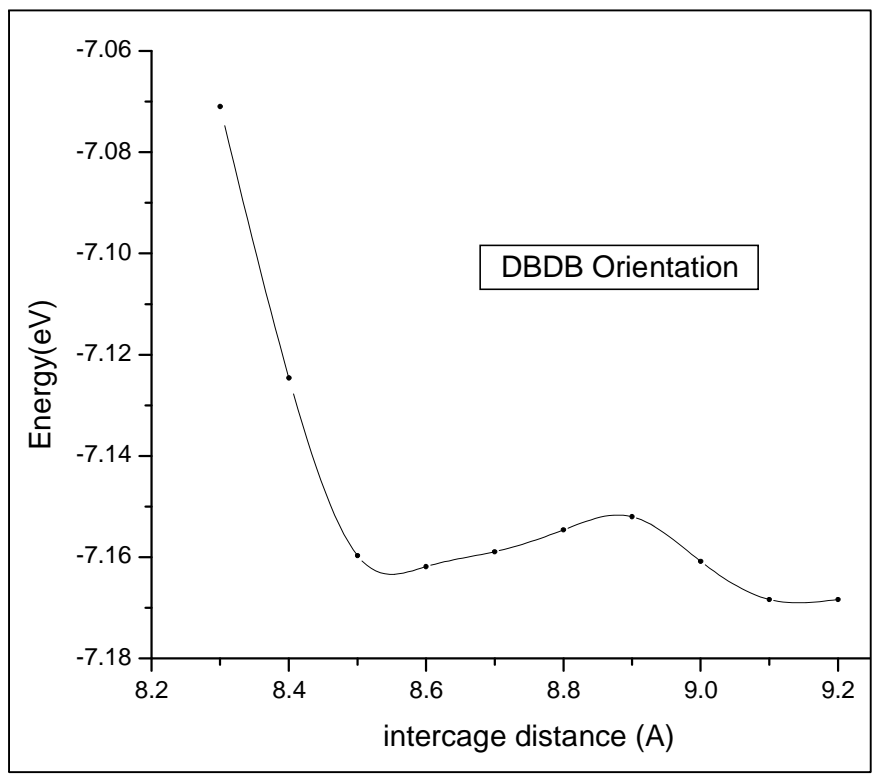

**Figure 2**

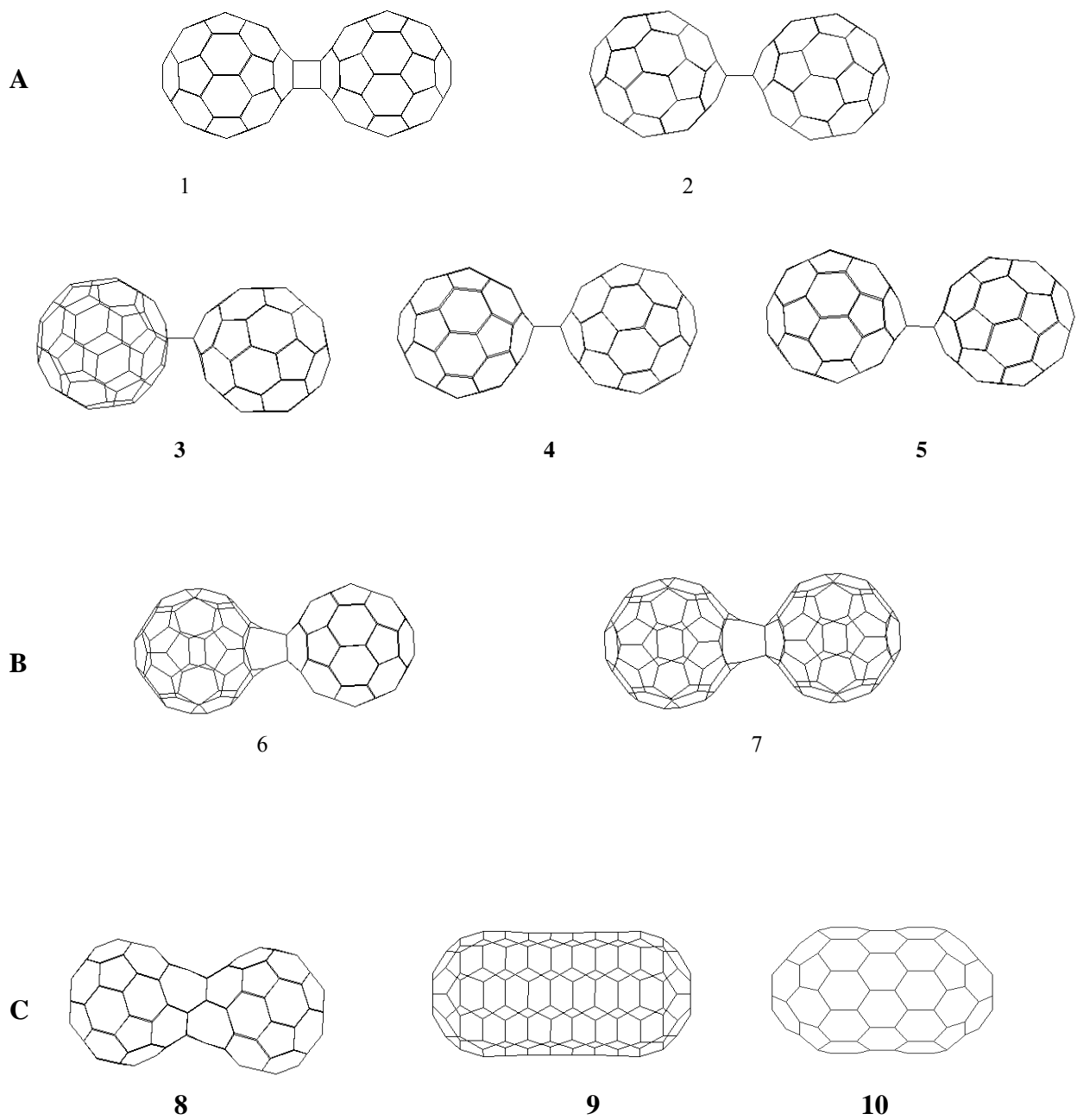

**Figure 3**

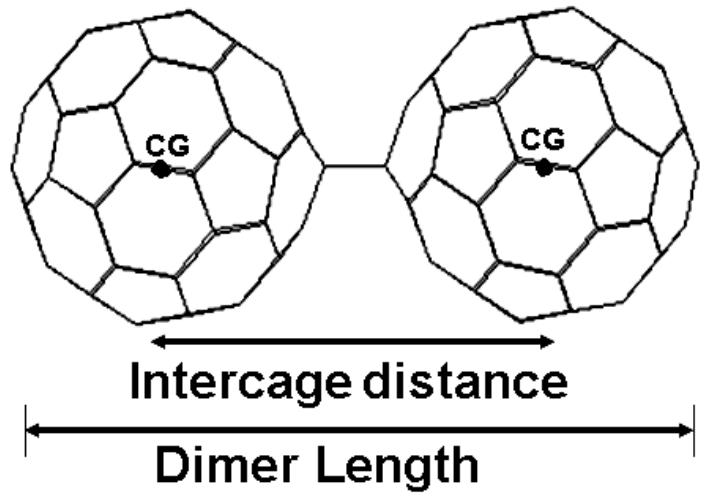

**Figure 4**

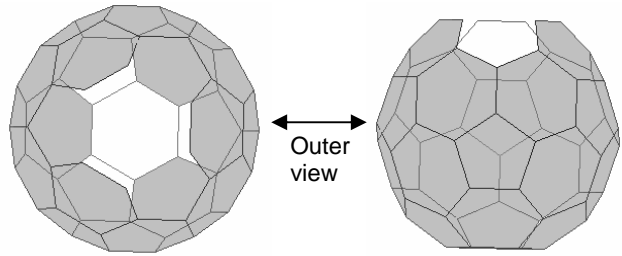

**Figure 5**

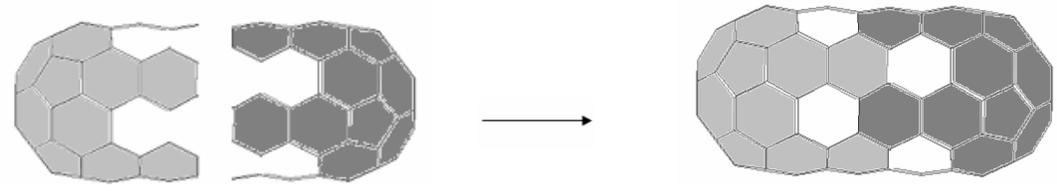

## Table Captions

**Table I:** Parameters of the potential.

**Table II:** The orientation and minimized energies of the ten $C_{60}$ dimer structures.

**Table III:** Recommended pressure and temperatures for the ten structures. The pressures correspond to initial required inter-cage distance, d.

**Table I**

| Tersoff Parameters | Original (Tersoff, 1998 b) | Modified |
|---|---|---|
| A(eV) | 1393.6 | 1380.0 |
| B(eV) | 346.7 | 349.491 |
| $\lambda_1$ (Å$^{-1}$) | 3.4879 | 3.5679 |
| $\lambda_2$ (Å$^{-1}$) | 2.2119 | 2.2564 |
| $\lambda_3$ (Å$^{-1}$) | 2.2119 | 2.2564 |
| $\beta$ | 1.57 x 10$^{-7}$ | 1.57 x 10$^{-7}$ |
| n | 0.72751 | 0.72751 |
| c | 38049. | 38049.0 |
| d | 4.3484 | 4.3484 |
| h | -0.57058 | -0.57058 |
| R (Å) | 1.95 | 1.95 |
| D (Å) | 0.15 | 0.15 |

**Table II**

| Structure No. | Structure as shown in Fig 3 | Starting Orientation | Initial Center to Center Dis. (Å) | Initial energy (eV) | Minimized energy (eV) | No. of inter cage bonds | Intercage Bond length (Å) | Final Center to Center Dis. (Å) | Dimer Length (Å) |
|---|---|---|---|---|---|---|---|---|---|
| 1 | Dumbell | DB - DB | 8.5 | .79 | -1.52 | 2 | 1.55 each | 8.89 | 15.94 |
| 2 | Single bonded-1 | C - C | 8.4 | .57 | -2.18 | 1 | 1.48 | 9.07 | 15.97 |
| 3 | Single bonded-2 | SB - PF | 8.5 | 2.67 | -1.48 | 1 | 1.51 | 8.81 | 15.84 |
| 4 | Single bonded-3 | DB - DB | 9.0 | 0 | -2.09 | 1 | 1.48 | 8.98 | 16.01 |
| 5 | Single bonded-4 | DB - PF | 8.5 | 3.86 | -2.04 | 1 | 1.48 | 9.02 | 16.1 |
| 6 | Fused – 2 | SB - DB | 8.5 | -.66 | -4.23 | 2 | 1.45 each | 8.91 | 15.97 |
| 7 | Fused – 3 | SB-SB | 8.5 | 2.06 | -4.30 | 2 | 1.44 each | 8.89 | 15.94 |
| 8 | Peanut | Open Hf - Closed Hf | 8.1 | 10.61 | -16.3 | 6 | 1.39 each | 8.50 | 15.44 |
| 9 | Armchair nano tube | Open PF - open PF | 8.0 | 130.08 | -33.64 | 6 | 1.41 each | 3.92 | 11.84 |
| 10 | Zigzag nanotube | *$C_{120}$ isomer | - | 26.93* | -35.10 | 6 | 1.41 each | 6.29 | 12.30 |

* Facing pentagons of two $C_{60}$ isomers are opened and these monomers are brought closer. We have minimized the $C_{120}$ isomer with our potential model, so initial energy required is much more than quoted here.

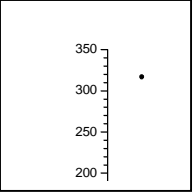

## Table III

| d(Å) | P (kBar) | T (K) | Structure | Category | *Velocity of $C_{60}$ mol. Beam $(ms^{-1})$* |
|---|---|---|---|---|---|
| 8.0 | 17.0 | 4135. | 9 | C-3 | 5901 |
| 8.1 | 17.0 | 4135. | 8 | C-1 | 1685 |
| 8.4 | 10.0 | 2600. | 2 | A-2 | 390 |
| 8.5 | 8.5 | 2230. | 1, 3, 5, 6, 7 | A-1, 3, 5 B- 1,2 | 460, 845, 1017 420, 743 |
| 9.0 | 4.0 | 1004. | 4 | A-4 | 0 |